\documentclass[%
aps,
prl,
reprint,
superscriptaddress,
nofootinbib,
amsmath,
amssymb
]{revtex4-2}
\usepackage{
	physics,
	graphicx,
	multirow,
	tabularx,
	mathtools, 
	placeins, 
	nicefrac, 
	hyperref, 
	threeparttable, 
	siunitx,
	enumitem,
	makecell,
	amsmath,amssymb,amsthm,mathrsfs,amsfonts,dsfont
}

\usepackage[english]{babel}

\usepackage[dvipsnames]{xcolor}
\hypersetup{
	colorlinks,
	linkcolor={blue!50!black},
	citecolor={blue!50!black},
	urlcolor={blue!80!black}
}

\usepackage[caption=false]{subfig} 
\begin{document}

\title{Light-induced trion-exciton competition revealed by ultrafast photoemission}

\author{Ittai~Sidilkover}
\author{Nir~Hen~Levin}
\affiliation{School of Physics and Astronomy, Faculty of Exact Sciences, Tel Aviv University, Tel-Aviv, 6997801, Israel}
\affiliation{Center for Light-Matter Interaction, Tel Aviv University, Tel-Aviv, 6997801, Israel}

\author{Yuval~Nitzav}
\affiliation{Department of Physics, Technion, Haifa, 3200003, Israel}

\author{Shiri~Gvishi}
\affiliation{School of Physics and Astronomy, Faculty of Exact Sciences, Tel Aviv University, Tel-Aviv, 6997801, Israel}
\affiliation{Center for Light-Matter Interaction, Tel Aviv University, Tel-Aviv, 6997801, Israel}

\author{Abigail~Dishi}
\affiliation{Department of Physics, Technion, Haifa, 3200003, Israel}

\author{Shaked~Rosenstein}
\affiliation{School of Physics and Astronomy, Faculty of Exact Sciences, Tel Aviv University, Tel-Aviv, 6997801, Israel}
\affiliation{Center for Light-Matter Interaction, Tel Aviv University, Tel-Aviv, 6997801, Israel}

\author{Noam~Ophir}
\affiliation{Department of Physics, Technion, Haifa, 3200003, Israel}

\author{Irena~Feldman}
\affiliation{Department of Physics, Technion, Haifa, 3200003, Israel}
\affiliation{The Helen Diller Quantum Center, Technion, Haifa, 3200003, Israel}

\author{Andrei~Varykhalov}
\affiliation{Helmholtz-Zentrum Berlin für Materialien und Energie, Elektronenspeicherring BESSY II, Albert-Einstein-Str. 15, 12489 Berlin, Germany.}

\author{Naaman~Amer}
\affiliation{School of Physics and Astronomy, Faculty of Exact Sciences, Tel Aviv University, Tel-Aviv, 6997801, Israel}
\affiliation{Center for Light-Matter Interaction, Tel Aviv University, Tel-Aviv, 6997801, Israel}

\author{Amit~Kanigel}
\author{Anna~Keselman}
\affiliation{Department of Physics, Technion, Haifa, 3200003, Israel}
\affiliation{The Helen Diller Quantum Center, Technion, Haifa, 3200003, Israel}

\author{Iliya~Esin}
\affiliation{Department of Physics, Bar-Ilan University, 52900, Ramat Gan, Israel}

\author{Hadas~Soifer}
\email[Corresponding author:]{hadassoifer@tauex.tau.ac.il}
\affiliation{School of Physics and Astronomy, Faculty of Exact Sciences, Tel Aviv University, Tel-Aviv, 6997801, Israel}
\affiliation{Center for Light-Matter Interaction, Tel Aviv University, Tel-Aviv, 6997801, Israel}

\begin{abstract}
Strong Coulomb interactions in low-dimensional quantum materials give rise to emergent bound states such as excitons and trions. Trions are conventionally secondary excitations, requiring both optical excitation and charge doping. In quasi-one-dimensional Ta$_2$NiS$_5$, however, an exceptionally large binding energy exceeding the single-particle band gap stabilizes an equilibrium trion gas under surface doping alone. Here, using time- and angle-resolved photoemission spectroscopy, we show that trions can also be generated purely optically, without external charge. Following photoexcitation of pristine Ta$_2$NiS$_5$ we observe a bright, momentum-localized, in-gap feature with a slow, fluence-dependent relaxation. Rate-equation modeling identifies it as a mixed population of trions and excitons, with the trions formed via an unconventional single-particle pathway that requires no photoexcited hole. The trion-exciton composition is controlled by pump fluence, with trions dominating the late-time relaxation. In surface doped samples, the same model with unchanged rates reproduces the full dynamics, linking the light-induced trions to their equilibrium counterparts through pump-induced dissociation and recapture. These results establish trARPES as a direct probe of charged quasiparticles far from equilibrium and open routes to optical control of neutral and charged excitations in correlated materials. 
\end{abstract}

\maketitle

In one-dimensional electronic systems, Coulomb interactions are amplified by spatial confinement and reduced dielectric screening, leading to the formation of excitons - tightly bound electron-hole pairs \cite{wannier_structure_1937,scholes_excitons_2006,anantharaman_excitonphotonics_2021}. The large binding energies in such systems facilitate excitons as robust quasiparticles that can couple to additional degrees of freedom, drive the emergence of a gapped excitonic insulator (EI) phase \cite{jerome_excitonic_1967,kaneko_new_2025,kunes_excitonic_2015}, or form higher-order complexes such as trions \cite{bondarev_relative_2014}, which are charged quasiparticles containing an exciton and an additional electron or hole. Angle-resolved photoemission spectroscopy (ARPES) \cite{sobota_angle-resolved_2021,gedik_photoemission_2017} has revealed these phenomena in quasi-one-dimensional crystals \cite{smolenski_large_2025,ma_angle_2025,liebich_controlling_2025,klein_bulk_2023}, including exciton-polaron interplay in $\text{Ta}\text{Se}_3$ \cite{ma_multiple_2022}, band flattening in the EI phase of $\text{Ta}_2\text{Pd}_3\text{Te}_5$\cite{huang_evidence_2024,zhang_spontaneous_2024}, and most recently, the observation of an equilibrium trion gas in Ta$_2$NiS$_5$ upon surface electron dosing \cite{nitzav_trion_2025}. 

Time- and angle-resolved photoemission spectroscopy (trARPES) \cite{boschini_time-resolved_2024} has emerged as a powerful technique for resolving light-induced dynamics across the energy-momentum landscape of excitonic systems \cite{volckaert_external_2023,bange_probing_2024}. It has enabled direct access to excitonic states in two-dimensional materials \cite{reutzel_probing_2024,madeo_directly_2020,dong_direct_2021,gosetti_unveiling_2025,karni_through_2023,tanimura_surface_2023} and led to the discovery of excitonic states at the surface of a topological insulator \cite{mori_spin-polarized_2023,mori_possible_2025}. By disentangling the interplay between multiple degrees of freedom \cite{mori_spin-polarized_2023,kunin_momentum-resolved_2023,christiansen_theory_2019,karni_structure_2022}, trARPES directly reveals exciton formation and decay pathways. An analogous dynamical picture for trions is being explored theoretically \cite{meneghini_arpes_2026,wu_what_2025}; however, it has remained out of experimental reach.

Here, we investigate optically driven electron dynamics in Ta$_2$NiS$_5$ using trARPES and report the observation of transient trions generated by an unconventional mechanism that does not require external charge doping. The trions emerge within a bright in-gap feature whose spectral properties match those of the equilibrium trion gas, and whose fluence-dependent dynamics are quantitatively captured by coupled rate equations describing competing formation channels: a single-particle pathway producing trions and a two-particle pathway producing excitons. Together, these identify the feature as a mixed population of trions and excitons, whose composition is controlled by pump fluence and which becomes trion-dominated at late times.
Finally, photoexciting a sample that hosts the equilibrium trion gas, we find that the same model, with unchanged rates and a single additional capture channel, reproduces the full dynamics, therefore directly linking the light-induced trions to their stable counterparts. 

\begin{figure*}
	\includegraphics[width=1\textwidth]{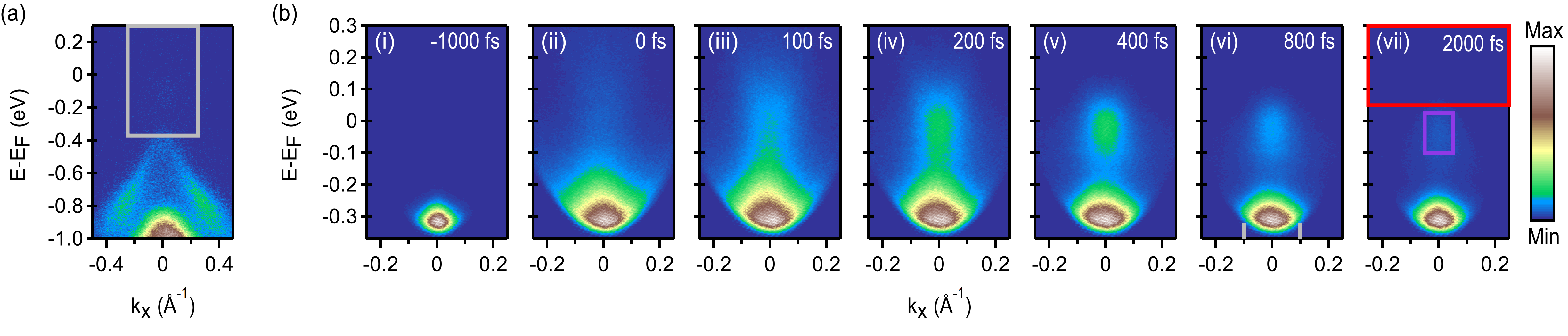}
    \caption{
        Time-resolved ARPES of Ta$_2$NiS$_5$.
		(a) ARPES spectrum obtained with $43$~eV photons. The gray rectangle marks the momentum and energy regions shown in each panel of (b). (b) Time-resolved ARPES spectra of Ta$_2$NiS$_5$ oriented to $\Gamma-X$ direction and pumped with a $800$~nm ultrafast pulse ($160$~$\mu$J/cm$^{2}$). The parabolic cutoff at the bottom of each panel is the photoemission horizon of the $6$~eV probe.
        }
	\label{fig:1_snapshots}
\end{figure*}

We examine light-induced dynamics in a freshly cleaved Ta$_2$NiS$_5$ sample. Although sharing quasi-one-dimensional structure with Ta$_2$NiSe$_5$ \cite{wakisaka_excitonic_2009,baldini_spontaneous_2023}, photoemission experiments establish it as a direct-gap semiconductor [see Fig.~\ref{fig:1_snapshots}(a) and refs \cite{chen_anomalous_2023,nitzav_trion_2025}]. Here we use 6~eV probe photon energy, which limits access to the very top of the valence band (VB), confined within the parabolic photoemission horizon. Figure~\ref{fig:1_snapshots}(b) shows trARPES spectra over a range of pump-probe delays, within the region marked by the gray rectangle in (a).
Upon excitation with the pump pulse (800~nm), the conduction band (CB) becomes populated, and the gap is completely filled by a narrow-in-momentum electronic distribution. At later times, as the excited carriers relax, a bright feature emerges within the gap where the unpumped system hosts no states. This feature reaches maximum intensity at $400$~fs, well after the 50~fs pump pulse has ended, and decays on a timescale of $2$~ps.

\begin{figure}
	\includegraphics[width=0.47\textwidth]{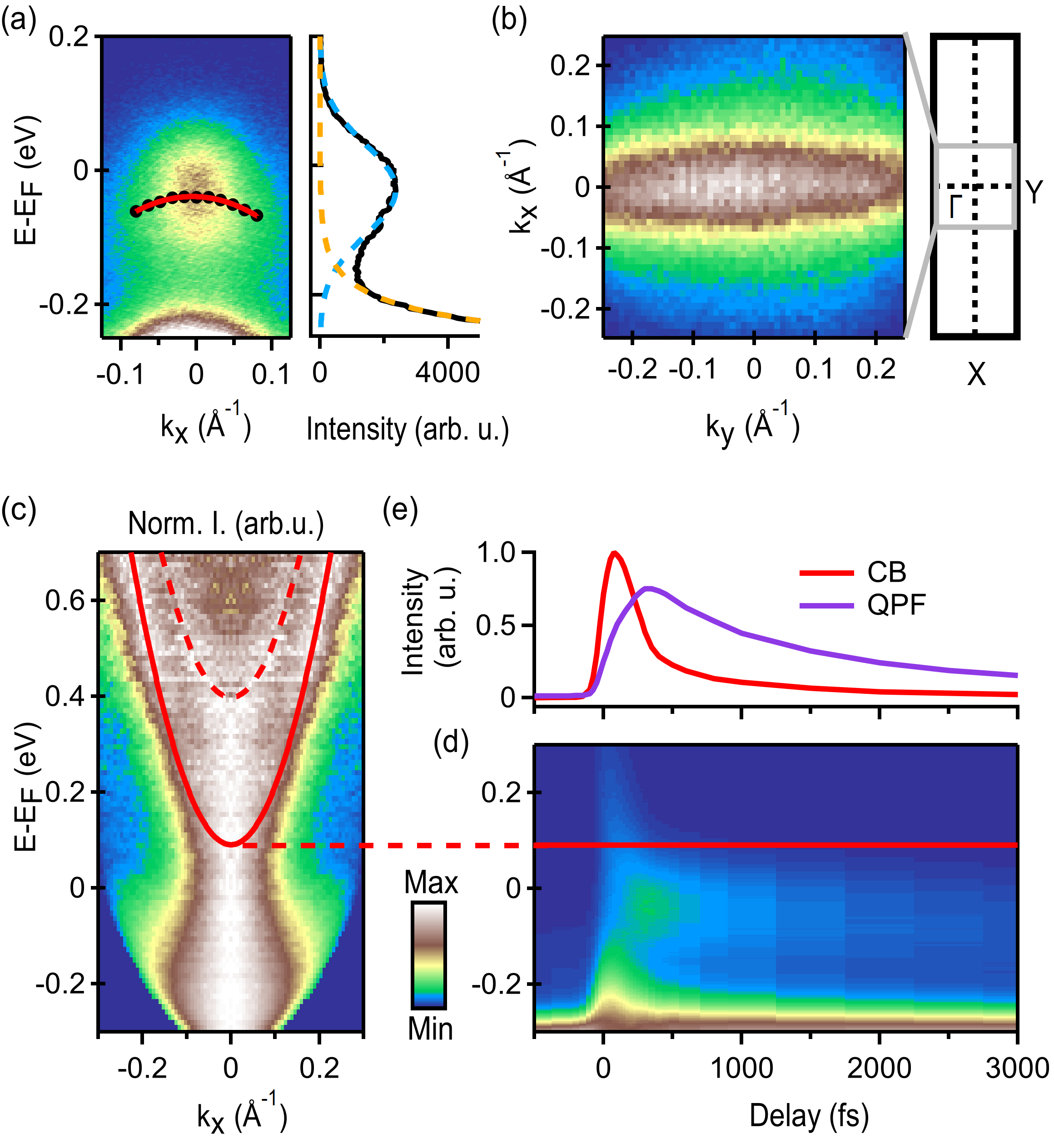} 
    \caption{
        Spectral and temporal characteristics of the QPF. (a) Dispersion of the QPF at $t=800$~fs (red), obtained by fitting the QPF peaks (black markers). The peaks are extracted by fitting the EDC (right panel, black) with a Gaussian to capture the QPF (light blue), and an exponent to capture the background (orange) \cite{SM}. (b) Fermi surface (FS) map of the QPF, taken at $t=400$~fs using the deflection mode of the electron analyzer (left) and a sketch of the captured portion of the Brillouin zone (right).
        (c) trARPES spectrum at $t=0$, obtained with a $400$~nm pump pulse, and normalized by MDC maximum for each energy bin. The extended energy range encompasses the two conduction bands. Red curves show the fitted band dispersions \cite{SM}. (d) EDC as a function of pump-probe delay, integrated over the momentum region between the gray markers in Fig.\ref{fig:1_snapshots}(b-vi) for each delay. The red line marks the fitted bottom of the CB. (e) Intensity vs delay of the CB and the quasiparticle feature integrated over the $k_x$ and $E$ ranges defined by the red and purple regions in Fig.\ref{fig:1_snapshots}(b-vii), respectively.}
    \label{fig:2}
\end{figure}

In Figure~\ref{fig:2} we examine the spectral and dynamical properties of this bright in-gap feature, which we denote as the quasiparticle feature (QPF).
First, the dispersion of the QPF [extracted at $t=800$~fs, see Fig.~\ref{fig:2}(a)] is hole-like, with an effective mass of $-0.96\pm0.03~m_e$. The hole-like dispersion is a clear indication that the QPF is not electron accumulation at the bottom of the CB, and is typically taken as a hallmark for excitons \cite{mori_spin-polarized_2023,gosetti_unveiling_2025}. However, the effective mass is much larger than the VB mass ($\sim0.3~m_e$ \cite{nitzav_trion_2025}) which is the expected value for excitonic features in trARPES spectra \cite{rustagi_photoemission_2018}. Second, the QPF center binding energy is well below the Fermi level; an anomaly for a neutral exciton, whose spontaneous formation would then be energetically favorable, destabilizing the ground state (EI instability \cite{jerome_excitonic_1967,kaneko_new_2025}). 
Lastly, the QPF momentum distribution is highly anisotropic, as shown in the Fermi surface in Fig.~\ref{fig:2}(b). While highly localized along $\Gamma-X$, the QPF extends the entire span of the Brillouin zone along $\Gamma-Y$. This would be unusual for excitons, which are typically highly localized in both in-plane momenta, forming a pocket below the band minimum.

Recent works suggest that trions can play a significant role in strongly correlated quasi-one-dimensional materials \cite{ma_multiple_2022,katoch_giant_2018,nitzav_trion_2025,ophir_trion_2026}. Particularly in Ta$_2$NiS$_5$, equilibrium ARPES measurements show that surface dosing establishes an equilibrium trion gas caused by excess electron density \cite{nitzav_trion_2025}. Calculations show that the total trion binding energy 
exceeds the single-particle band gap, making it energetically favorable for a (doping-induced) CB electron to transition into a trion state. Indeed, its dispersion is comparable with predictions for trion photoemission \cite{meneghini_arpes_2026}. The
momentum anisotropy of the QPF transient is similar to that measured for the equilibrium trion gas, which likewise resides below the Fermi level, though the light-induced feature appears at a higher energy (lower binding energy) than its equilibrium counterpart (see SM \cite{SM}). 

These results place trions as the leading candidates for the QPF and suggest that photoexcited electrons undergo a process analogous to the doping-induced one. Conventionally, trion formation is a secondary process: a photoexcited electron and hole first bind into an exciton, which subsequently captures a second CB electron \cite{mak_tightly_2013,nakama_trion_2024}. In Ta$_2$NiS$_5$, the energetics permit a more direct route \cite{nitzav_trion_2025}: a single CB electron can convert into a trion by generating an electron-hole pair out of the filled valence band and binding to it. The binding energy gain exceeds the pair-creation cost, with the difference released to phonons \cite{trovatello_ultrafast_2020}. The formation rate therefore depends on the CB electron density alone, $\propto n_e$, in contrast to the two-particle exciton channel, which also requires a photoexcited hole, $\propto n_e n_h$ \cite{ciocys_manipulating_2020,peters_carrier_2019}. 
Since it is energetically favorable, the trionic band is expected to serve as a major decay path for excited electrons, forming an effective band minimum. Figure~\ref{fig:2}(c) highlights the CB in a normalized spectrum at $t=0$~fs, revealing two parabolic, single-particle, conduction bands, with the minimum located 90~meV above the Fermi level \cite{SM}. The normalization further emphasizes the complete filling of the gap at $t=0$ with $k$-localized states. At later times, the QPF emerges below the CB bottom simultaneously with the complete evacuation of the latter [Fig.\ref{fig:2}(d)], confirming that trion formation is a major relaxation channel for CB electrons. This is further emphasized in Fig.\ref{fig:2}(e), which plots integrated intensities versus delay, indicating the QPF intensity reaches a substantial portion of the peak intensity observed over the entire CB. 

Two significant differences exist between the doping- and light-induced processes \cite{nguyen_equilibrium_2025}. First, in the light-induced scenario both electrons and holes are introduced by the pump, enabling exciton as well as trion creation -- unlike the doped case at equilibrium, where the charge imbalance allows only for trion creation. Second, the light-induced trions are unstable and ultimately decay via interactions with photo-holes, keeping the system charge neutral at all times. This interaction can generally result in a pair of excitons, implying their constant presence in a system with light-induced trions. The QPF therefore comprises both types of quasiparticles, appearing as a single feature because their photoemission energies differ by only tens of meV which is below our energy resolution (see SM \cite{SM}).

\begin{figure*}

	\includegraphics[width=0.75\textwidth]{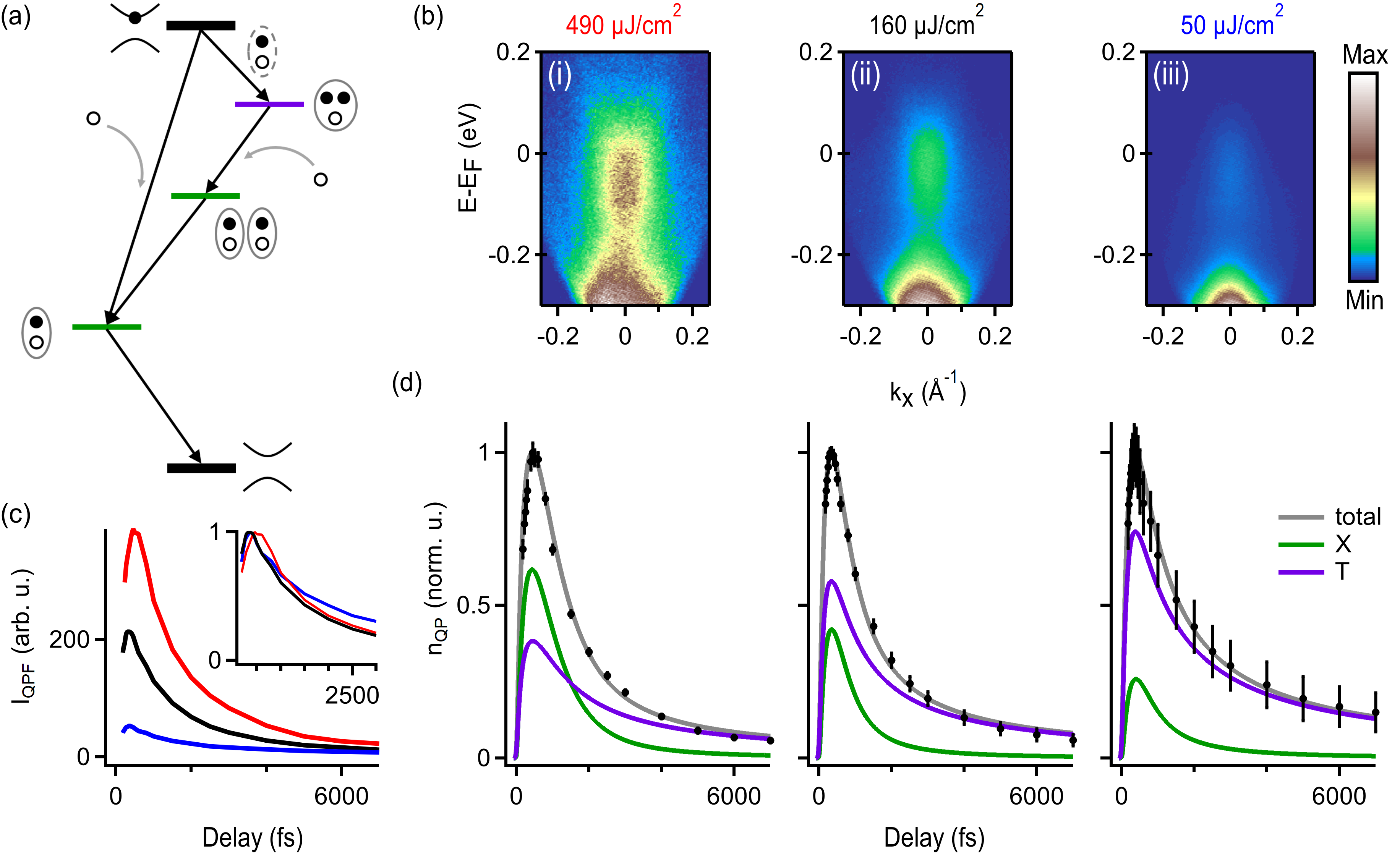} 
    \caption{QPF fluence dependence and time-dependent composition.
        (a) sketch of the processes described by the rate equations, and the quasiparticle states. The two-particle processes (exciton formation and trion decomposition) are marked with an additional gray arrow, indicating the interaction with a photo-induced hole. Dashed oval: electron-hole pair generated from the filled VB during trion formation.
        (b) trARPES snapshots at $t=300$~fs for three pump fluences ($800$~nm): $490$, $160$ and $50$~$\mu$J/cm$^{2}$, respectively. The color scale is set to the minimum and maximum of each panel separately. 
        (c) QPF intensity of the three fluences, with normalized curves in the inset.  
        (d) Normalized QPF intensity and the total quasiparticle population, $n_{X}+n_{T}$, from the fit results shown in black and gray, respectively. Exciton ($n_X$) and trion ($n_T$) populations are shown in green and purple, respectively. 
        }
    \label{fig:3}
\end{figure*}

A schematic of the quasiparticle creation and decay processes in this system is sketched in Fig.~\ref{fig:3}(a). Trion creation is a single-particle process with a rate proportional to $n_e$, whereas (non-resonant) exciton creation requires both photo-induced electrons and holes, with a rate proportional to $n_e n_h$.
The decay channels invert this pattern: excitons decay via spontaneous electron-hole recombination, a one-particle process depending only on the exciton population ($n_X$), whereas trion decay is a two-particle process scaling with the trion and hole populations ($n_T n_h$). This distinction between one- and two-particle processes provides a means to disentangle their respective contributions through the QPF dynamics. Since both $n_e$ and $n_h$ scale linearly with excitation fluence to first order, the exciton creation rate is expected to increase quadratically with fluence, whereas the trion creation rate should exhibit a linear dependence.

We study the dependence of the QPF on pump fluence in Fig.~\ref{fig:3}(b-d). Throughout, the QPF is quantified by fitting the energy distribution curves (EDCs) to a Gaussian peak on an exponential background accounting for the VB tail [Fig.~\ref{fig:2}(a); \cite{SM}]. Panel (b) shows spectra measured at $t=300$~fs for three pump fluences spanning an order of magnitude. Beyond the trivial increase in brightness, the dynamics exhibit a strong fluence dependence, as seen in Fig.~\ref{fig:3}(c) with changes in rise time, peak steepness and decay times across fluences. To explore these non-trivial trends, we construct a set of rate equations describing the creation and decay of excitons and trions following photoexcitation:

\begin{equation}
    \label{rate_eq}
\begin{split}
    \frac{dn_{e}}{dt}&=-C_{T}\ n_{e}-C_{X}\ n_{e} n_{h}\\
    \frac{dn_{h}}{dt}&=-C_{X}\ n_{e} n_{h}-R_{T}\ n_{T} n_{h}\\
    \frac{dn_{X}}{dt}&=C_{X}\ n_{e} n_{h}+2R_{T}\ n_{T} n_{h}-R_{X}\ n_{X}\\
    \frac{dn_{T}}{dt}&=C_{T}\ n_{e}-R_{T}\ n_{T} n_{h},
\end{split}
\end{equation}
where $C_T$ ($C_X$) and $R_T$ ($R_X$) denote the trion (exciton) creation and decay rates, respectively. We simplify the model by assuming equal rates for the single-particle processes, $C_T = R_X \equiv \alpha_{\mathrm{1p}}$, and for the two-particle processes, $C_X = R_T \equiv \alpha_{\mathrm{2p}}$. The former equates trion formation and exciton decay: both involve an exciton that spontaneously forms from or decays into the vacuum, and therefore have comparable rates. The latter equates processes requiring interaction between a hole and either a trion or an electron; since both carry the same negative charge, the attractive interaction is expected to be similar (see SM \cite{SM} for details). We restrict the model to single-photon processes and neglect higher-order contributions (like the aforementioned secondary trion creation).

We fit this model to data at three fluences to extract the two rates for each case (End Matter).  Fig.~\ref{fig:3}(d) compares the experimental QPF intensity (black) with the fitted total quasiparticle population, $n_T(t)+n_X(t)$ (gray), demonstrating excellent agreement. The fit further disentangles the individual trion and exciton populations, revealing a strong fluence dependence of their ratios. 
While trions dominate at late times ($t > 2000$~fs) across all fluences, excitons dominate at early times at high fluence but remain a minority at intermediate and low fluences. This behavior is consistent with the predicted linear and quadratic fluence dependences of the two channels.

The fitted rates are summarized in Table~\ref{tab:rate_eq}. Notably, both effective rates decrease systematically with increasing fluence, slowing by a factor of 2-4 over the order-of-magnitude fluence range. The common direction and comparable magnitude of these trends point to a shared origin, which we attribute to screening of Coulomb interactions. The rates in Eq.~\eqref{rate_eq} are thus understood as effective constants at a given excitation density. Beyond this overall slowdown, the dynamics within each fluence follow the predicted density dependence: as $n_e$ decreases with time, both formation channels decelerate, but the two-particle processes, additionally dependent on the depleting hole density $n_h$, decelerate more rapidly. Trions therfore dominate at late times, as both their decay and exciton formation become increasingly suppressed, with a small exciton population persisting as a decay product of the trions. This is further supported by the asymptotic behavior of the rate-equation solution, which follows a power law with $n_X\propto t^{-2}$ and $n_T\propto t^{-1}$ (see SM).

We note that QPF dynamics alone do not select the model; an exciton-only description reproduces the measured intensity across all three fluences (End Matter). It does so, however, only by sustaining exciton formation from a residual free-carrier population lasting for picoseconds, much slower than the observed CB evacuation [Fig.~\ref{fig:2}(d-e)]. 

\begin{figure}
    \centering
	\includegraphics[width=0.47\textwidth]{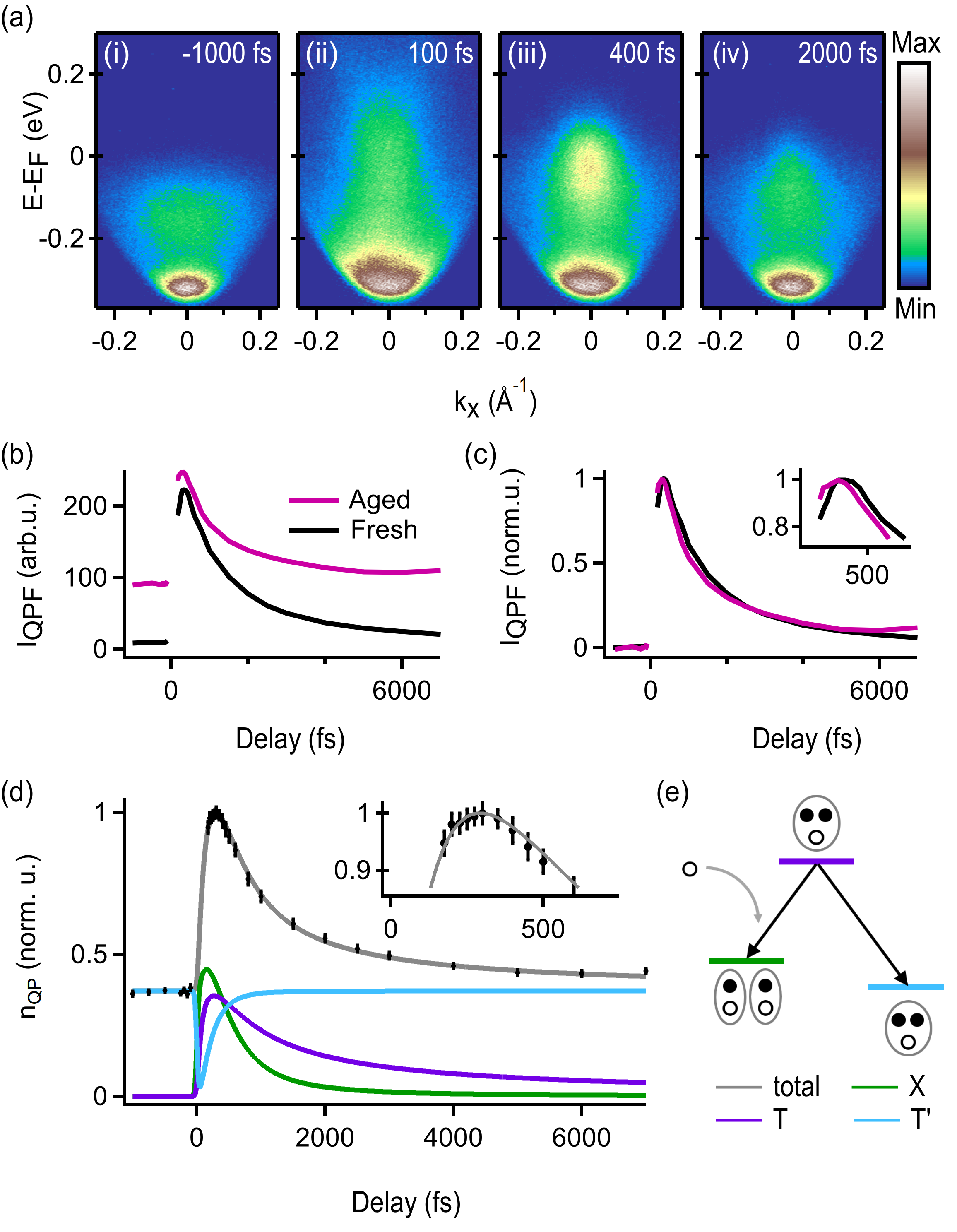} 
    \caption{Dynamics of trions in an aged sample.
        (a) Time-resolved ARPES spectra of an aged sample with $800$~nm ultrafast pulse ($160$~$\mu$J/cm$^{2}$). (b) total QPF intensity of fresh (black) and aged (magenta) samples. (c) same as (b) but normalized to one, following the subtraction of pre-pump intensity for the aged sample. Inset: early times. (d) QPF integrated intensity and the fitted  total quasiparticle population, $n_X+n_T+n_T'$, shown in black and gray, respectively. Exciton ($n_X$), excited trion ($n_T$), and ground-state trion ($n_{T'} $) populations in green, purple, and light blue, respectively. (e) sketch of trion states showing both excited and ground-state trions within the band gap.}
    \label{fig:4}
\end{figure}

The dynamics discussed so far involve trions created entirely by the pump. To probe their relation to the equilibrium trions, we exploit the gradual electron doping of the cleaved surface as it ages in vacuum, which establishes the equilibrium trion gas characterized in Ref.~\cite{nitzav_trion_2025}. In the aged sample studied here, this gas is directly visible as a pre-pump in-gap population of $n_{0}=0.37$ of the peak QPF intensity [Fig.~\ref{fig:4}(a)]. When photoexcited with $160$~$\mu$J/cm$^{2}$ [same as Fig.~\ref{fig:3}(b-ii)], the aged sample reproduces the early-time hot-electron distribution of the fresh sample above the Fermi level [see SM], while its QPF peak is only $\sim10\%$ brighter, despite the added equilibrium population. The QPF dynamics differ in both regimes [Fig.~\ref{fig:4}(b-c)]: the initial rise is faster than in the fresh sample, and the late-time decay is slower, with the QPF remaining well above the fresh-sample response throughout the measured range and decaying toward the equilibrium baseline.

These observations are captured by a minimal extension of the fresh-sample model, with a single additional process. The pump dissociates the equilibrium trions into excitons and free electrons. The injected excitons bypass the quasiparticle formation bottleneck, accounting for the faster rise. Newly formed excited trions are progressively captured back into the equilibrium doping-induced trion states, where, lacking the internal energy required for hole-assisted breakup, they are removed from the decay channel. The QPF decay therefore slows as the ground-state reservoir refills, producing the slow decay rate. 
The fit results of this model (End Matter) are shown in Figure~\ref{fig:4}(d). With the creation and decay rates held fixed at the fresh-sample values, the capture rate is the only free dynamical parameter, yielding $\alpha_{\mathrm{c}}^{-1}=212\pm{46}$~fs, comparable to the trion formation time (237$\pm$24~fs). The remaining fitted quantities set only the intensity scale: the overall normalization and the equilibrium density $n_0$, the latter tightly constrained by the measured pre-pump baseline.
The fit returns an initial electron density equal to the fresh-sample value, in agreement with the measured hot-electron distributions.

The late-time asymptote is due to charge conservation: the excess electron density guarantees that, once the photo-induced holes are depleted, a trion population equal to $n_{0}$ survives. The dynamics before it distinguish between the models: those without the protected reservoir decay to the baseline too rapidly, whereas models with inert equilibrium trions inherit the slower fresh-sample rise time (see End matter for model comparison). The entire aged-sample dataset is thus captured by the fresh-sample rates plus a single capture channel. Aging alters the initial trion population, not the underlying dynamics.

In summary, we have shown that trions can form in Ta$_2$NiS$_5$ upon photoexcitation alone without charge doping, through a single-particle channel. The trion state acts as an emergent conduction-band minimum and the dominant relaxation pathway for photoexcited electrons, rapidly depleting the CB. This channel competes with conventional two-particle exciton formation, and the balance between them is set by the excitation density, making the composition of the bound state population optically tunable. Finally, the same dynamics link the light-induced trions to the equilibrium trion gas: in surface-doped samples, the pump dissociates the equilibrium trions and the photoexcited electrons are recaptured into the emptied ground states. These results establish quasi-one-dimensional Ta$_2$NiS$_5$ as a model system for the dynamics of complex bound states in strongly interacting, low-dimensional materials, and demonstrate that trARPES gives direct access to charged many-body quasiparticles far from equilibrium.  

\section{acknowledgments}
We thank Roni Ilan and Tobias Holder for fruitful discussions. 
We thank Helmholtz-Zentrum Berlin für Materialien und Energie for the allocation of synchrotron radiation beamtime.
I.S. acknowledges funding from the National Quantum Science and Technology Program of the Israeli Planning and Budgeting Committee.
H.S. acknowledges the support of the Zuckerman STEM Leadership Program, the Young Faculty Award from the National Quantum Science and Technology program of the Israeli Planning and Budgeting Committee, the ERC PhotoTopoCurrent 101078232 and the Israel Science Foundation (Grant No. 2117/20).
A.K. acknowledges funding by the Israel Science Foundation (Grant No. 2443/22).

\bibliography{paper_bib_with_urls}

\section*{End Matter}

\textit{Appendix A - details on rate equations and initial conditions}:  
The rate equations follow the dynamics of photo-excited electrons and holes, tracing the formation and decay of excitons and trions. The initial condition reflects the effect of the ultrafast pump pulse, which promotes electrons into the CB and leaves the same number of holes in the valence band. To avoid numerical instabilities and erroneous favoring of the two-particle process, the hard initial condition $n_e(t=0)=n_h(t=0)=const$ is replaced with a smoothed introduction of holes and electrons, over a timescale representing the pump pulse duration:

\begin{equation}
    \label{rate_eq_intial_condition}
\begin{split}
    \frac{dn_{e}}{dt}&=-C_{t}n_{e}-C_{x}n_{e}n_{h} + pF\,S(t)\\
    \frac{dn_{h}}{dt}&=-C_{x}n_{e}n_{h}-R_{t}n_{t}n_{h} + pF\,S(t).\\
\end{split}
\end{equation}
With $S(t)=\frac{1}{\sqrt{2\pi}\Delta_{t}}e^{-\frac{(t-3\Delta_{t})^2}{2\Delta_{t}^2}}$ acting as a source of electrons and holes at early stages of the numerical calculation. The Gaussian shape of $S$ follows the envelope of a generic ultrafast pulse. The value of $\Delta_{t}$ was set to $25$~fs, to match the experimental setup. The prefactor is set to be the fluence $F$ multiplied by a constant $p$ for converting from fluence to the scale of particle number. The value of $p$ was fitted globally to keep the fluence hierarchy between sets (single value for all three fluences), the same value was later used for the analysis of the aged sample (see below). The experimental data is always normalized to 1, with the corresponding intensity normalization of the model given by the fit. For the fresh sample data, the small (and unavoidable) equilibrium trion signal is subtracted prior to normalization. The time axis of the simulation is shifted after the calculation concludes by $3\Delta_{t}=75$~fs such that $t=0$ coincides with the peak of $S(t)$, matching the convention of the experimental data. Only delay values $\geq175$~fs were accounted for in the fit to avoid earlier delays when the system still thermalizes after the pump pulse, and where the QPF parameters are not reliably fitted across all fluences.

\begin{table}
	\centering
	\begin{tabular}{lcccr} 
		\\
		\hline
		Fluence ($\mu$J/cm$^2$) & $\alpha_{1p}$ (ps$^{-1}$) & $\alpha_{2p}$ (ps$^{-1}$) & $\alpha^{-1}_{1p}$ (fs) & $\alpha^{-1}_{2p}$ (fs)\\
		\hline
	    50 (low)&5.23$\pm$1.36&6.82$\pm$1.73&191$\pm$49&146$\pm$37\\
		160 (medium)&4.22$\pm$0.42&3.87$\pm$0.40&237$\pm$24&258$\pm$27\\
		490 (high)&2.21$\pm$0.16&1.54$\pm$0.11&452$\pm$33&649$\pm$46\\
		\hline
	\end{tabular}
    	\caption{Fitting values for rate equation parameters.
		Values for the rates obtained from fitting the three data sets (fluences) presented in Fig.~\ref{fig:3}, with $\alpha_{2p}$ quoted per unit normalized density.}
            	\label{tab:rate_eq}
\end{table}

\begin{figure}
    \centering
    \includegraphics[width=1\linewidth]{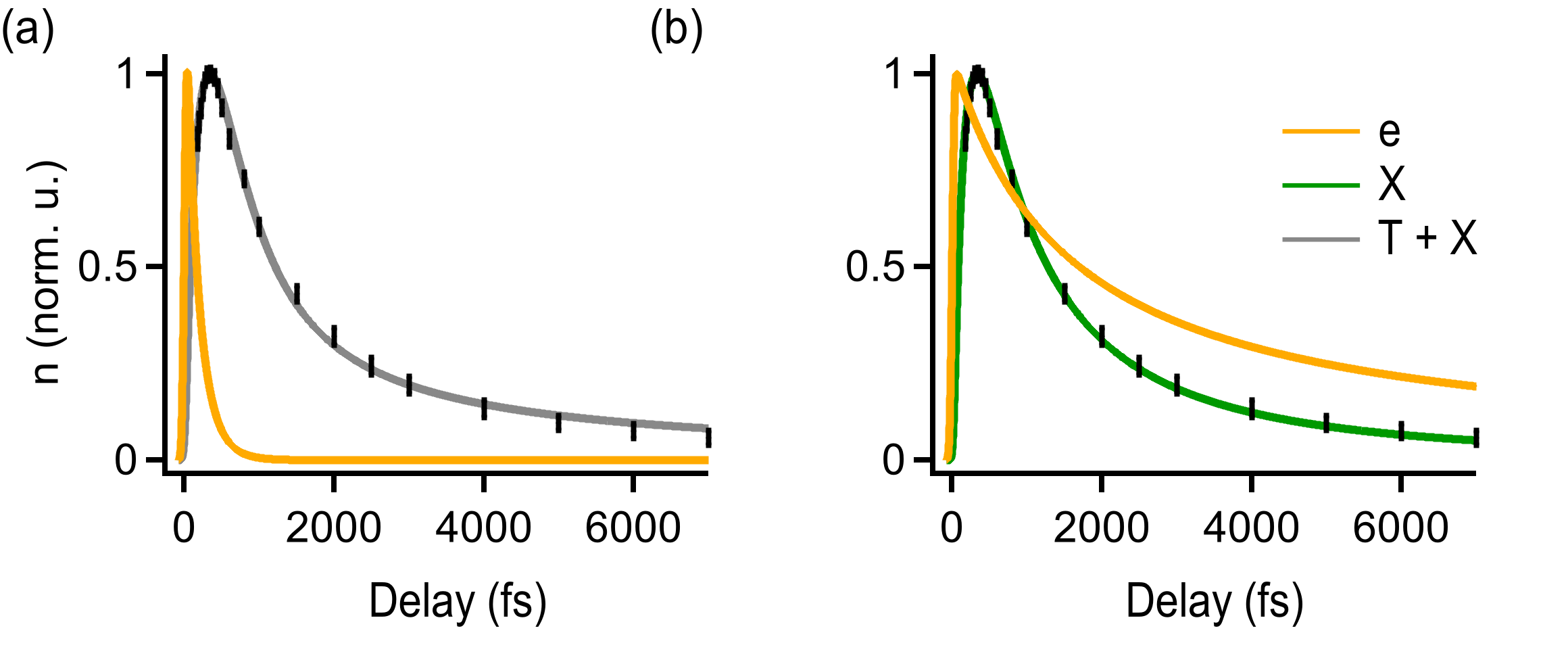}
    \caption{Comparison to exciton only model. Black dots are QPF integrated intensity for the mid-fluence dataset. (a) Full model fit: $n_X+n_T$ shown in gray. (b) Exciton only model: $n_X$ shown in green. Normalized electron populations $n_e$ of both models shown in orange. 
    }
    \label{fig:em_exciton_model}
\end{figure}

\textit{Appendix B - Exciton-only model for the fresh sample}:
We test whether the fresh-sample dynamics can be described without a trion population, using a single bound species with two-particle formation and one-particle decay: $dn_X/dt=C_X n_e n_h - R_X n_X$. This model reproduces the measured QPF intensity [Fig.~\ref{fig:em_exciton_model}(b), shown for 160~$\mu$J/cm$^2$; the behavior at the other fluences is equivalent]. However, the model sustains the slow, late-time QPF only through continuous exciton formation, requiring electron and hole populations that persist throughout the measured delay window, in contradiction to the observed electron dynamics [Fig.~\ref{fig:2}(e)]. In the full model, by contrast, the single-particle trion channel drains the CB completely [Fig.~\ref{fig:em_exciton_model}(a)], and no residual electron population remains. The two models are therefore distinguished by the free-carrier population they predict. 
Including an exciton-exciton annihilation channel, $-Bn_X^2$, does not alter this picture: the fit converges to the same electron-sustained solution, with negligible contribution from this annihilation term. More generally, any description restricted to neutral excitons, regardless of its decay kinetics, is excluded by the doped sample (next appendix): its equilibrium in-gap population at negative delays and its finite late-time baseline, protected by charge conservation, are natural for a charged quasiparticle and unavailable to a neutral one. 

\textit{Appendix C - Rate-equation model for the aged sample}: 
The aged sample is described by a minimal extension of the fresh-sample model, Eq.~\eqref{rate_eq}. The pump dissociates the equilibrium trion population of density $n_{0}$ into excitons and free electrons.
A single term is added, describing phonon-mediated re-capture of excited trions ($T$) into the trion ground states ($T'$) that were emptied by the pump:
\begin{equation}
    \label{eq:em_reservoir}
\begin{split}
    \frac{dn_{T}}{dt}&=C_{T}\,n_{e}-R_{T}\,n_{T}n_{h}
    -\alpha_{\mathrm{c}}\,n_{T}\!\left(1-\frac{n_{T'}}{n_{0}}\right)\\
    \frac{dn_{T'}}{dt}&=\alpha_{\mathrm{c}}\,n_{T}\!\left(1-\frac{n_{T'}}{n_{0}}\right),
\end{split}
\end{equation}
with the $n_{e}$, $n_{h}$, and $n_{X}$ equations unmodified, except for the different pump-induced initial conditions. 
The pump-induced dissociation is introduced through the same smooth source (appendix A): $-n_{0}S(t)$ is added to the trion equation and $+n_{0}S(t)$ to the electron and exciton equations, so the total prefactor of S in the $n_e$ equation reads $1+n_0$. 

\begin{figure*}
    \centering
    \includegraphics[width=0.75\textwidth]{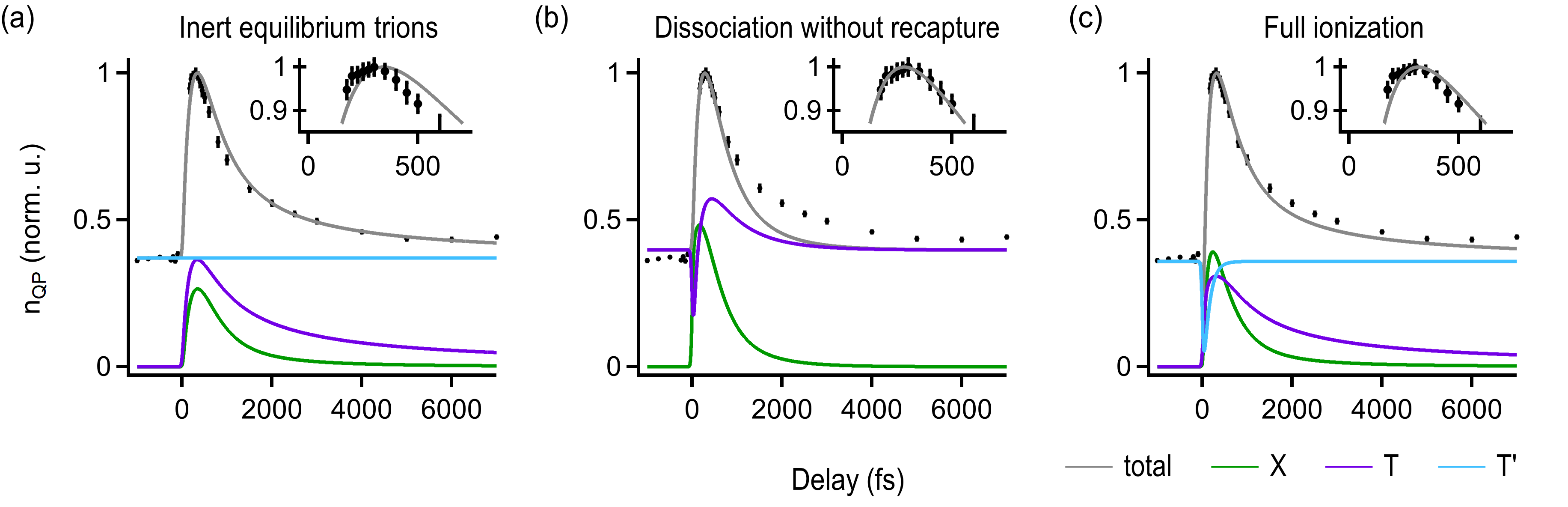}
    \caption{Alternative aged sample models.
    Fitting results (gray) of the QPF intensity of the aged sample [same as Fig.\ref{fig:4}(d)] and quasiparticle populations of exciton (green), excited trions (purple), and ground-state trions (light blue) for three alternative models. (a) Inert equilibrium trions, (b) Dissociation of equilibrium trions without recapture to a ground state, and (c) Full ionization of the trions into electrons and holes during excitation. 
    }
    \label{fig:em_model_comparison}
\end{figure*}

Captured trions occupy the trion ground states associated with the surface doping; lacking the internal energy required for hole-assisted breakup, they no longer participate in the decay channel. The number of such states is fixed by the charge imbalance, which sets the capacity $n_{0}$ in the vacancy factor $(1-n_{T'}/n_{0})$; capture therefore self-limits as the reservoir refills, and the ground state population cannot exceed its equilibrium value. The measured QPF intensity is compared with the total in-gap population, $n_{X}+n_{T}+n_{T'}$.

The fit is maximally constrained: $\alpha_{\mathrm{1p}}$ and $\alpha_{\mathrm{2p}}$ are held fixed at the fresh-sample values for the same fluence ($160$~$\mu$J/cm$^{2}$), yielding the capture rate  $\alpha_{\mathrm{c}}^{-1}=212\pm46$~fs. As in the fresh-sample fits, the overall intensity normalization is fitted, and here the equilibrium density $n_0$ therefore also enters as a fit parameter, tightly constrained by the measured pre-pump in-gap intensity (0.37 of the QPF peak). The fitted value ($n_0=0.37\pm0.04$) is consistent with this measurement.

\textit{Appendix D - Comparison of alternative models for aged sample}: 
We have examined alternative descriptions of the equilibrium trions under photoexcitation, each constructed as a minimal modification of Eq.~\eqref{rate_eq} with the fresh-sample rates held fixed, and each removing a single ingredient of the full model. (i)~\emph{Inert equilibrium trions}: the equilibrium trions take no part in the dynamics, and the QPF is the fresh-sample solution atop a constant pedestal $n_{0}$. (ii)~\emph{Dissociation without recapture}: the equilibrium trions are dissociated as above, but no stable ground states are available ($\alpha_{\mathrm{c}}=0$). (iii)~\emph{Full ionization}: Upon pumping, the equilibrium trions completely dissociate into two electrons and a hole, with the ground-state reservoir retained; no bound in-gap complex survives the excitation.

A constraint common to all variants follows from charge conservation. The aged sample carries a fixed excess electron density given by the pre-pump trion density $n_0$. The photoinduced holes deplete via hole-assisted trion decay, and once they are exhausted, an in-gap population equal to the excess charge necessarily survives. All charge-conserving models therefore share the same late-time asymptote at the equilibrium baseline $n_{0}$. The models are distinguished solely by their dynamics. 
 
Figure~\ref{fig:em_model_comparison} compares the models with the measured aged-sample dynamics; each alternative fails where the removed ingredient acted. Model~(i) reproduces the slow tail but not the onset: the transient inherits the fresh-sample rise time, whereas the measured aged response is faster; 
Model~(ii) reproduces the fast rise due to the injected excitons but decays too rapidly: with no protected states, every trion remains exposed to hole-assisted breakup therefore accelerating the decay through the $-R_{T}n_{T} n_{h}$ term. 
Model~(iii) fails in both regimes ($\chi^{2}=6.2$, versus $0.7$ for the full model, $2.2$, and $17.2$ for models (i)-(ii), respectively): with no injected excitons the QPF rise is too slow, while the additional photo-released holes accelerate the late-time breakup; the fit moreover returns an ill-determined recapture rate, $\alpha_{\mathrm{c}}^{-1}=108\pm100$~fs, consistent with no capture at all, in contrast with the well-constrained and mutually consistent values obtained in the bound-complex variants.
Only the full model, Eq.~\eqref{eq:em_reservoir}, captures both the dissociation-injected excitons at early times, bypassing the formation bottleneck, and the slow late-time decay, produced by the progressive filling of the ground states, which removes trions from the breakup channel. 

\end{document}


\title{Supplementary Material for ``Light-induced trion-exciton competition revealed by ultrafast photoemission"}
\pagebreak

\maketitle

\tableofcontents
\clearpage

\section{Methods}
\subsection{Experimental methods}
Single crystals $\text{Ta}_2\text{Ni}\text{S}_5$ were synthesized via the chemical vapor transport (CVT) method. The quality of the samples was confirmed by XRD measurements, and the elemental composition was determined through energy-dispersive X-ray (EDX) analysis. 

Synchrotron-ARPES measurements [Fig.1(a)] were carried out at the $1^2$ instrument beamline at the BESSY II electron storage ring operated by the Helmholtz-Zentrum Berlin für Materialien und Energie, using $43$~eV photons. Samples were cleaved at $25$~K.

Time-resolved ARPES measurements were performed with 6~eV (206~nm) ultrashort probe pulses using the DA30-L electron analyzer (ScientaOmicron). The sample was kept at 80~K and at a pressure better than $1.5\ 10^{-10}$~mbar throughout the measurement. To increase the visible momentum range on the detector, a bias was applied between the sample stage and the entrance cone of the hemispherical analyzer \cite{pfau_low_2020, gauthier_expanding_2021}. The conversion from angle to momentum was carried out as described in \cite{gauthier_expanding_2021}. 
The 6~ev probe pulse was generated by fifth harmonic generation of a compressed 1030~nm pulse (Tangerine Short-pulse, Amplitude) using three BBO nonlinear crystals. Pump 800~nm 50~fs pulses were generated by an OPA system (Twinstarzz, Fastlite-Amplitude). 400~nm pulses were produced by second harmonic generation of the OPA signal. The overall time resolution is better than 85~fs for both pump wavelengths, as evaluated by cross-correlation at high energy bands. The energy resolution is $50$~meV.

\subsection{Photoemission from Excitons and Trions}
Photoemission of an electron comprising an exciton leaves behind a VB hole, yielding a kinetic energy reduced from that of a CB electron by the exciton binding energy $E_{bin}^X$, such that $E_\mathrm{k}=E_\mathrm{CB}-E_{bin}^X$ \cite{rustagi_photoemission_2018}. Alternatively, the photoemission process from a trion leaves behind a bound exciton, and the measured kinetic energy of the emitted electron becomes $E_\mathrm{k}=E_\mathrm{CB}-E_{bin}^{X-e}$ below the CB, where $E_{bin}^{X-e}$ denotes the additional binding energy of an electron to an exciton \cite{wu_what_2025,nitzav_trion_2025,meneghini_arpes_2026}. Signals from excitons and trions are therefore expected at different energies; however, both will appear inside the gap, and the difference between the two is expected to be on the order of tens of meV~\cite{nitzav_trion_2025}, comparable to the observed QPF width and the energy resolution, which hinders their spectral separation.

\subsection{QPF fitting}
\begin{figure}
	\centering
	\includegraphics[width=0.75\textwidth]{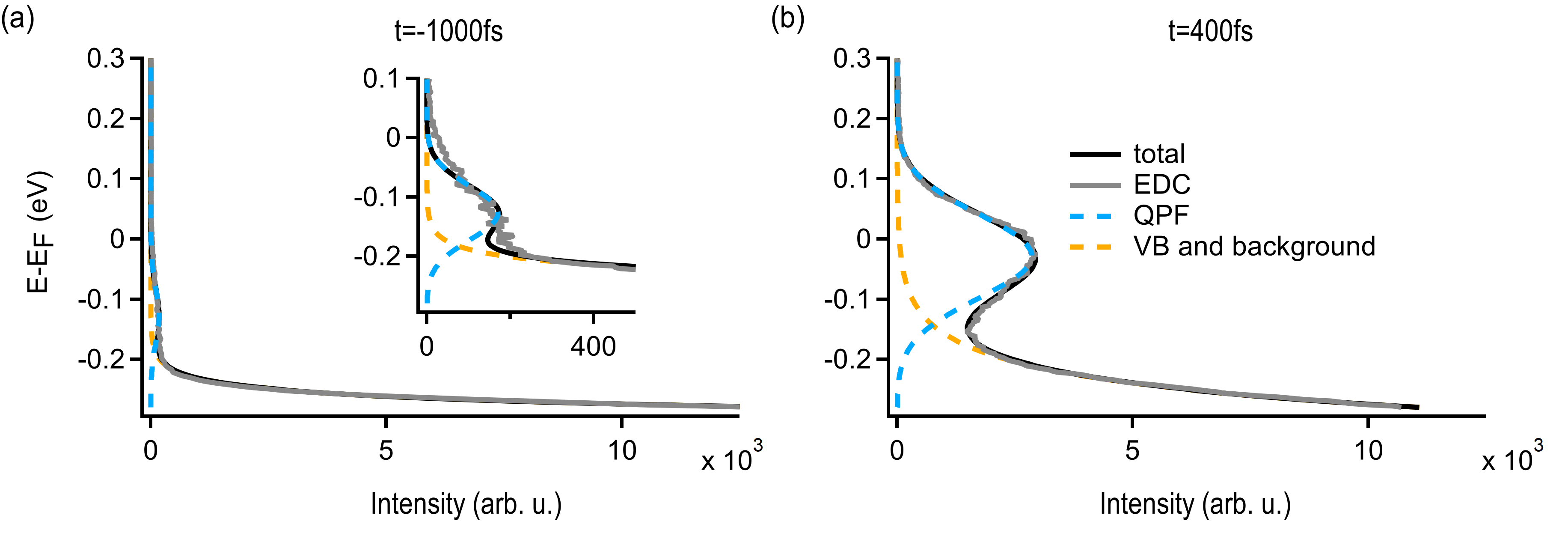} 
	\caption{Fitting of the QPF.
         (a) EDC at $t=-1000$fs of the spectrum shown in Fig.1(b-i), integrated over $[-0.025,0.025]\angstrom^{-1}$ (gray) and fitted (black) to a sum of a Gaussian and a decaying exponent [Eq.\eqref{eq:QPF_fit}] to capture the QPF and expanding VB, shown in orange and light-blue, respectively. (b) Same as (a) obtained for $t=400$fs, Fig.1(b-v). Note the inset reflecting the equilibrium spectrum, which exhibits a small population of stable trions.
         }
	\label{fig:s_EDC_fit}
\end{figure}

\begin{figure}
	\centering
	\includegraphics[width=1\textwidth]{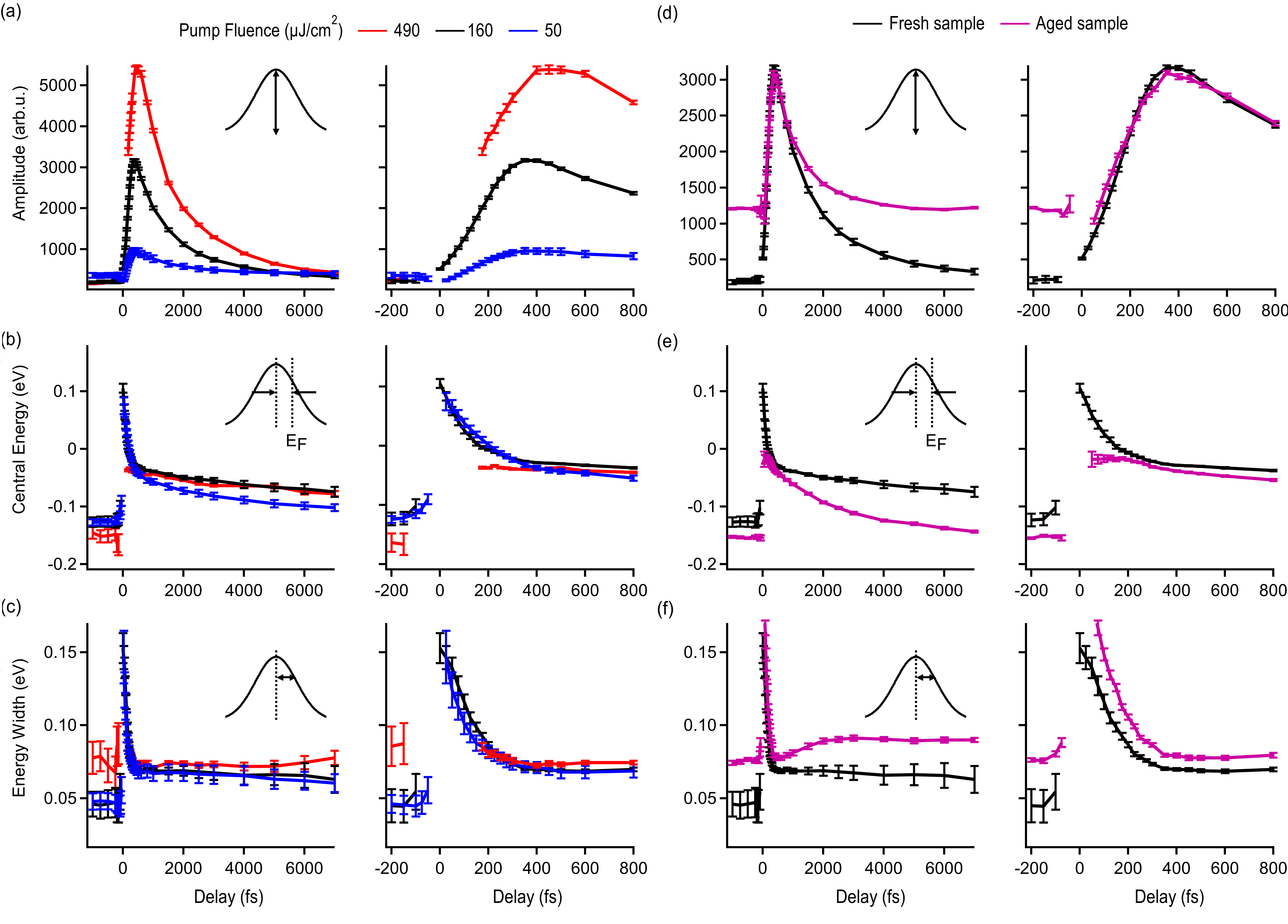} 
	\caption{
    QPF parameters. EDC fitting results of the QPF as a function of delay for the parameters defined in Eq.\eqref{eq:QPF_fit}: (a) Gaussian amplitude $A$, (b) Gaussian width $\sigma$, and (c) center position $E_0$ for three pump fluences presented in Fig.3 of the main text: high (red), medium (black) and low (blue). (d-f) same as (a-c) but for the fresh (black) and aged (magenta) samples. EDCs were integrated over $k_{x}=[-0.027,0.027]$~\text{\angstrom}$^{-1}$ and fitted over $E=[-280,280]$~meV. Points with unreliable fit were omitted. 
    }
	\label{fig:QPF_parameters}
\end{figure}

The total intensity of the QPF, used for the fitting of the rate equation in Fig.3 and Fig.4, is extracted from the fitting of an EDC capturing the QPF signal. An EDC spanning $E=[-0.28,0.28]$eV energy range is constructed by integrating over the counts in the momentum slice $k_{x}=[-0.027,0.027]$~\text{\angstrom}$^{-1}$, for each delay point. The resulting EDC is fitted with:
\begin{equation}
    I(E)=Ae^{-\frac{(E-E_0)^2}{2\sigma^2}}+Be^{-\frac{E}{C}},
    \label{eq:QPF_fit}
\end{equation}
where $Ae^{-\frac{E^2}{2\sigma^2}}$ captures the QPF parameters: the peak value $A$, its energy position $E_0$ and the width $\sigma$. The exponent $Be^{-\frac{E}{C}}$ captures the background, composed mainly of the valence band. The fitted total intensity is the product of the Gaussian amplitude and energy width. Exemplary fits, for both negative and positive delays are shown in Fig.\ref{fig:s_EDC_fit} and the parameters as a function of time are plotted in Fig.\ref{fig:QPF_parameters} for all the datasets. At early times, in addition to the quasiparticles, the fit captures the broad electron distribution, as evidenced by the values of $E_0$ and $\sigma$. The sharp early-time reduction of both can be attributed to the rapid cooling of the excited carriers that bind to form the QPF. At later times, after the quasiparticles have fully established, a slower downward trend of $E_0$ is observed on the picoseconds timescale. We attribute this change in binding energy to changes in dielectric screening induced by the coexistence of neutral and charged quasiparticles.    

\subsection{Aged sample}
\begin{figure}
	\centering
	\includegraphics[width=0.75\textwidth]{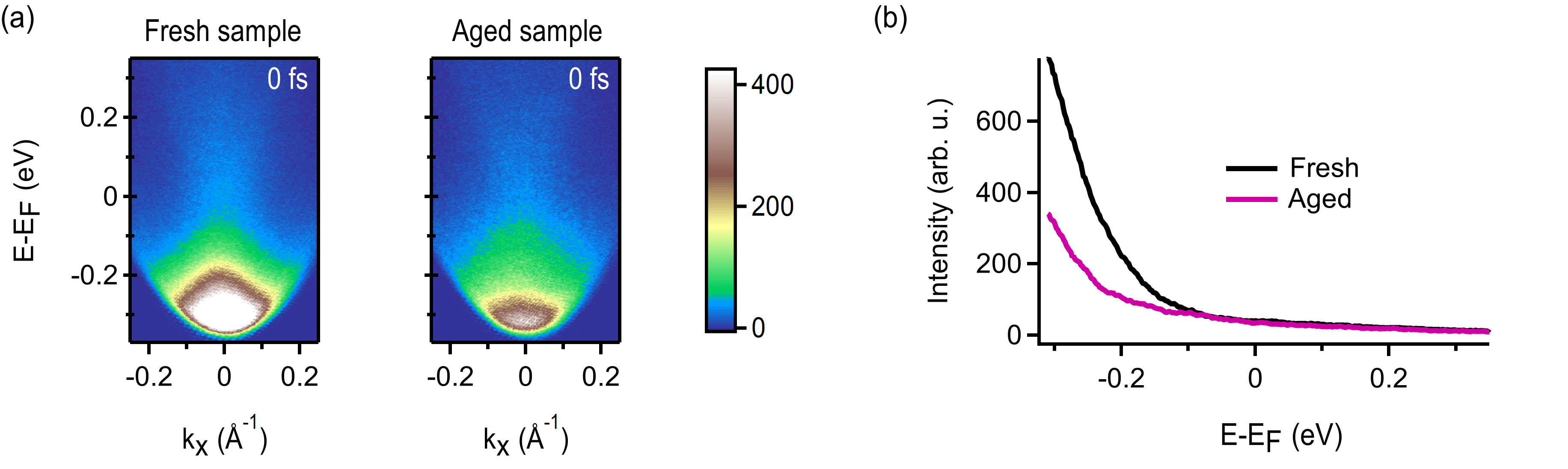} 
	\caption{Photoexcitation of fresh and aged sample. (a) The $t=0$~fs spectra of the fresh (left) and aged (right) sample are shown in the same color scale. (b) EDC averaged over $[-0.05,0.05]\angstrom^{-1}$ of the fresh sample (black) and aged (magenta) samples.}
	\label{fig:s_fresh_aged_EDC}
\end{figure}

The datasets shown in Fig.3 and Fig.4 were measured using the same cleaved sample. The measurements were performed several hours apart to allow the equilibrium trion population to develop under UHV conditions. The timescale of equilibrium trion buildup is consistent with Ref.\cite{nitzav_trion_2025}, which further observes a simultaneous downwards shift in energy of the VB. This downward shift is evident in the negative delay spectra shown in Fig.1(b-i) and Fig.4(a-i), which serve as equilibrium measurements. The equilibrium trion feature is directly visible in the fitted QPF parameters of the aged sample [Fig.~\ref{fig:QPF_parameters}(f)]: at negative delays, its center energy lies at $E_0=-160$~meV, below the Fermi level and $\sim80$~meV lower (higher binding energy) than the late-time energy of the light-induced feature. 
After aging, trARPES measurements were performed under the same pumping conditions as the middle fluence of the fresh sample. The $t=0$~fs spectra of both fresh and aged samples are presented in Fig.\ref{fig:s_fresh_aged_EDC}(a), set to the same color scale. The shifting of the VB is clearly visible across the intensity spread within the gap and above it. The hot-electron distributions from both measurements coincide perfectly for $E-E_F>-100$~meV, confirming identical pumping conditions.

\subsection{Conduction band analysis}
\begin{figure}
	\centering
	\includegraphics[width=0.75\textwidth]{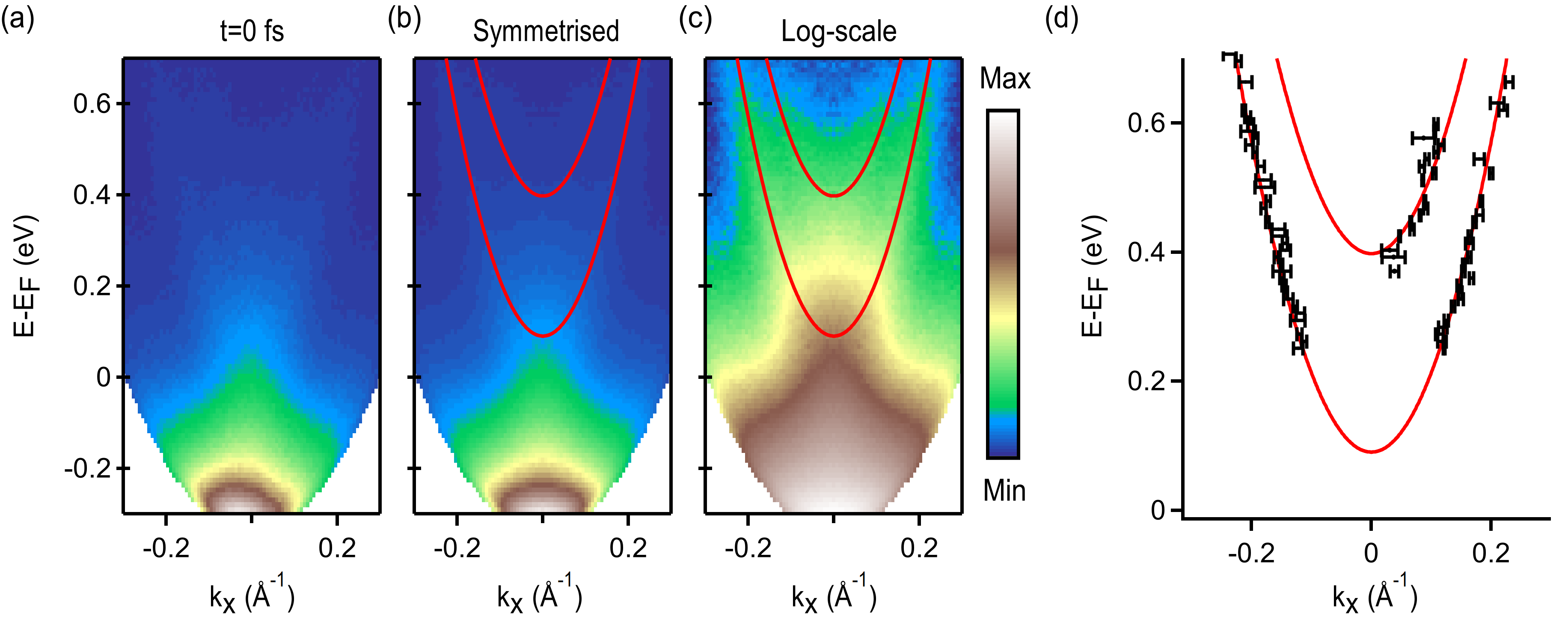}
	\caption{
    Conduction band analysis. (a) The raw spectrum of Fig.2(c), obtained at $t=0$~fs with $400$~nm pump. (b) Symmetrized spectrum overlaid with the fitted guidelines. (c) same as (b) in logarithmic scale. (d) bands peaks (black) extracted from fitting the MDCs of the inner and outer conduction bands, and the fit results of the band shape (red).
        }
	\label{fig:s_CB}
\end{figure}

We fit the dispersion of the two conduction bands, with the lower conduction band minimum at $90\pm30$~meV [red lines in Fig.2(c)]. 
The conduction-band dispersion were extracted from the symmetrized
spectrum at $t=0$ shown in Fig.\ref{fig:s_CB}(a). The band momenta were obtained from fitting the MDCs. The intensity is dominated by a hot-electron distribution spanning the gap and the conduction band, centered at $\Gamma$, with resolved conduction-band peaks at high energies appearing on the flanks of the central distribution. To disentangle them, the MDC is fitted by a sum of Gaussians. One, centered at $\Gamma$, along with a constant background, captures the central distributions, and one or two additional Gaussians to capture the CB peaks:
\begin{equation}
I(k) = b_0 + G_0\,\exp\!\left(-\frac{k^2}{2\sigma_0^2}\right)
      + \sum_{j} A_j\,\exp\!\left(-\frac{(k-k_j)^2}{2\sigma_j^2}\right).
\label{eq:mdc_model}
\end{equation}
Above $\approx 0.37$~eV, both bands are resolved, and two shoulder Gaussians were fitted, yielding the inner- and outer-band momenta simultaneously. Below this energy, only the outer band remains distinguishable; there, a single shoulder Gaussian was fitted, extending the outer-band dispersion down to $0.25$~eV. The outer (lower) conduction band peaks were extracted from the non-symmetrized data, while those of the inner (upper) band were extracted from the symmetrized spectrum, where the noise level is reduced. Unreliable MDC peaks were omitted. The two-band guidelines in Fig.\ref{fig:s_CB}(b) are fits of the MDC peaks. The outer band momenta peaks were fitted by:
\begin{equation}
    E(k)_{\mathrm{outer}} = E_0 + A\,\bigl[\,1 - \cos(c\,k)\,\bigr],
    \label{eq:cosine}
\end{equation}
The fit of the inner band was performed by fixing the band shape using the outer band parameters and fitting a constant energy shift: $E_{\mathrm{inner}}(k) = \Delta E+ E_{\mathrm{outer}}(k)$

\subsection{QPF dispersion}
\begin{figure}
	\centering
	\includegraphics[width=0.75\textwidth]{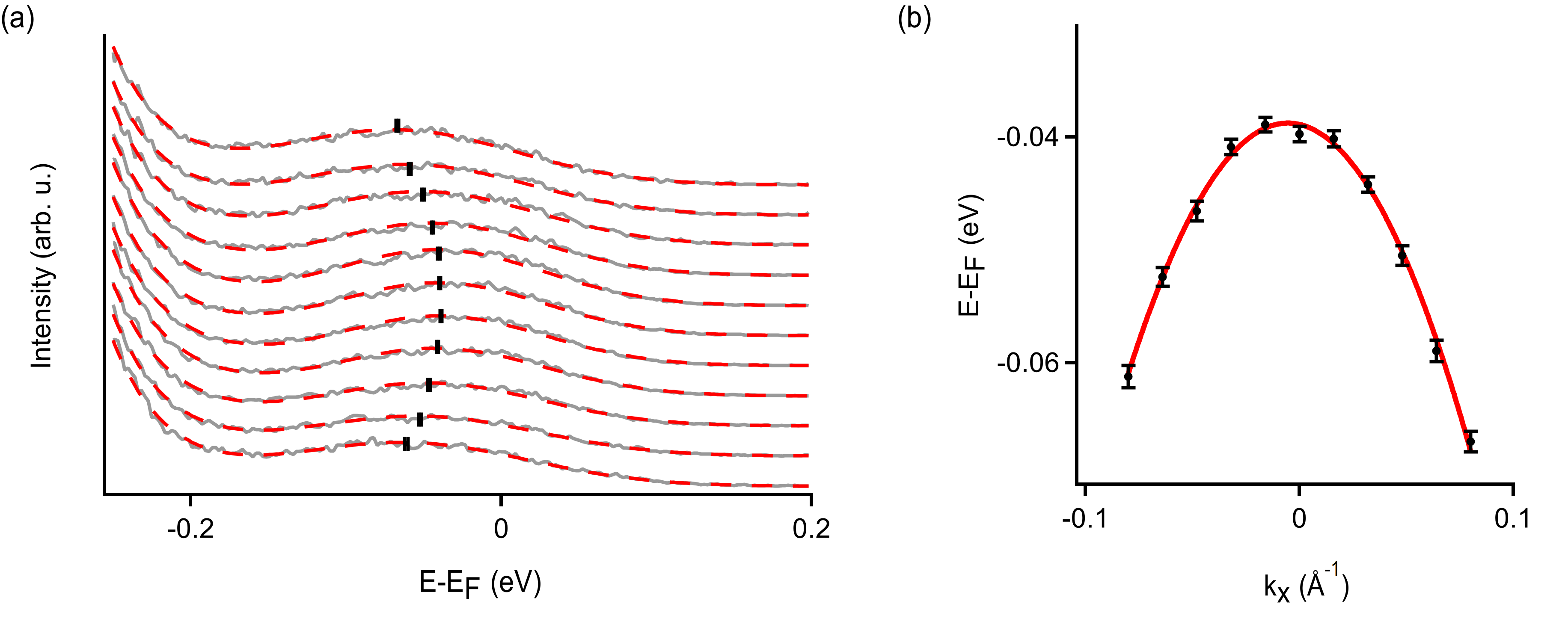} 
	\caption{
    QPF mass. (a) EDCs (gray) of 11 momentum slices and their fit (red). The QPF (Gaussian) peaks are shown in black. (b) parabolic dispersion of the QPF (red), obtained by fitting the EDC peaks, exhibiting a hole-like effective mass. 
    }
	\label{fig:mass}
\end{figure}

Figure~\ref{fig:mass}(a) shows EDCs for pump-probe delay of $800$~fs, integrating over 11 consecutive momentum regions. The EDCs are fitted in the same way as presented in Fig.\ref{fig:s_EDC_fit} -- including both VB and background as well as the QPF contributions -- with the QPF peaks shown with black markers and also plotted over the spectrum in Fig.2(a). The peak positions were fitted with a parabolic function, and the effective mass of $m^*=-0.96\pm0.03~m_e$ was extracted, as can be seen in Fig.~\ref{fig:mass}(b). 

\clearpage

\section{Rate equations and QP analysis}

\subsection{Rate equations parameters}
Here, we estimate the exciton creation and annihilation rates used in the rate equations. We assume that these processes are dominated by electron-phonon scattering, described by an electron-phonon interaction Hamiltonian. Specifically, we consider a local electron-phonon coupling to an optical phonon mode at zero momentum, with the interaction Hamiltonian given by
\begin{equation}
H_{\mathrm{ep}}=\int d^3\mathbf{r}\,\Big[g_{cc}\bigl(c_{\mathbf{r}}^\dagger c_{\mathbf{r}}-h_{\mathbf{r}}^\dagger h_{\mathbf{r}}\bigr)+g_{cv}\bigl(c_{\mathbf{r}}^\dagger h_{\mathbf{r}}^\dagger+c_{\mathbf{r}}h_{\mathbf{r}}\bigr)\Big](b^\dagger+b).
\end{equation}
Here, $c_{\mathbf{r}}^\dagger$ and $h_{\mathbf{r}}^\dagger$ create an electron in the conduction band and a hole in the valence band, respectively, while $b^\dagger$ creates an optical phonon. The coupling constants
\begin{equation}
g_{\alpha\beta}=\sum_i\left(\frac{a\hbar}{2M_iV\omega}\right)^{1/2}\mathbf{e}_i\cdot\mathbf{D}_{\alpha\beta}^i
\end{equation}
describe intraband ($\alpha\beta=cc$) and interband ($\alpha\beta=cv$) electron-phonon scattering processes, respectively. The index $i$ labels the ion position within the unit cell. In Ta$_2$NiS$_5$, the interband phonon coupling is expected to be significant due to the strong pseudo Jahn - Teller effect. The transition rates are obtained using Fermi’s golden rule and can be written as
\begin{equation}
\begin{aligned}
R_X &\propto \frac{2\pi}{\hbar}\frac{1}{V}
\int d^3\mathbf{r}_1\,
\left|\langle X(\mathbf{r}_1)|H_{\mathrm{ep}}|\phi\rangle\right|^2
\delta(E_X-E_\phi), \\[0.5em]
C_T &\propto \frac{2\pi}{\hbar}\frac{1}{V}
\int d^6\mathbf{r}_{1,2}\,
\left|\langle e(\mathbf{r}_1)|H_{\mathrm{ep}}|T(\mathbf{r}_2),\phi\rangle\right|^2
\delta(E_e-E_T-E_\phi), \\[0.5em]
C_X &\propto \frac{2\pi}{\hbar}\frac{1}{V^2}
\int d^9\mathbf{r}_{1,2,3}\,
\left|\langle e(\mathbf{r}_1)h(\mathbf{r}_2)|H_{\mathrm{ep}}|X(\mathbf{r}_3),\phi\rangle\right|^2
\delta(E_{eh}-E_X-E_\phi), \\[0.5em]
R_T &\propto \frac{2\pi}{\hbar}\frac{1}{V^2}
\int d^{12}\mathbf{r}_{1,2,3,4}\,
\left|\langle T(\mathbf{r}_1)h(\mathbf{r}_2)|H_{\mathrm{ep}}|X(\mathbf{r}_3)X(\mathbf{r}_4),\phi\rangle\right|^2
\delta(E_T+E_h-2E_X-E_\phi),
\end{aligned}
\end{equation}
where
\begin{equation}
|X(\mathbf{r})\rangle=\int d^3\mathbf{a}\,\Phi_X(\mathbf{a})\,c_{\mathbf{r}+\mathbf{a}/2}^\dagger h_{\mathbf{r}-\mathbf{a}/2}^\dagger|0\rangle
\end{equation}
and
\begin{equation}
|T(\mathbf{r})\rangle=\int d^6\mathbf{a}_{1,2}\,\Phi_T(\mathbf{a}_1,\mathbf{a}_2)\,c_{\mathbf{r}+\mathbf{a}_1-\mathbf{a}_2/3}^\dagger c_{\mathbf{r}-\mathbf{a}_1+\mathbf{a}_2}^\dagger h_{\mathbf{r}-\mathbf{a}_1-\mathbf{a}_2/3}^\dagger|0\rangle
\end{equation}

denote the excitonic and trionic wavefunctions, respectively, and $|\phi\rangle$ denotes the phonon state. Evaluating these expressions explicitly yields
\begin{equation}
\begin{aligned}
R_X &\propto \frac{2\pi}{\hbar} g_{cv}^2 \, |\Phi_X(0)|^2, \\[0.5em]
C_T &\propto \frac{2\pi}{\hbar} g_{cv}^2
\int d^3\mathbf{r}\,
\left|\Phi_T(\mathbf{r},0)-\Phi_T(0,\mathbf{r})\right|^2, \\[0.5em]
C_X &\propto \frac{2\pi}{\hbar} \left(g_{cc}^2+g_{vv}^2\right), \\[0.5em]
R_T &\propto \frac{2\pi}{\hbar} \left(g_{cc}^2+g_{vv}^2\right)
\int d^3\mathbf{r}_1\, d^3\mathbf{r}_2\,
\left|2\Phi_T(\mathbf{r}_1,3\mathbf{r}_2)\Phi_X(2\mathbf{r}_1)\Phi_X(2\mathbf{r}_2)\right|^2 .
\end{aligned}
\end{equation}

We note that the intraband and interband electron-phonon coupling constants can be parametrically different. In contrast, the geometric overlap factors associated with the excitonic and trionic wavefunctions are typically of order unity. As a result, we expect the rates $R_X$ and $C_T$ to be comparable (both $\propto g_{cv}^2$), and similarly expect $C_X$ to be comparable to $R_T$ (both $\propto \left(g_{cc}^2+g_{vv}^2\right)$).

\subsection*{Asymptotic late-times behavior}
At late times, we expect the electron population to be nearly depleted, as the trion-creation channel dominates the early-time dynamics. In this limit, namely for $n_e = 0$, the rate equations reduce to
\begin{align}
\frac{dn_h}{dt} &= - R_T n_T n_h, \nonumber \\
\frac{dn_T}{dt} &= - R_T n_T n_h, \nonumber \\
\frac{dn_X}{dt} &= 2 R_T n_T n_h - R_X n_X .
\label{eq:rate_equations}
\end{align}

Using charge neutrality, which once $n_e = 0$ is simply given by $n_T = n_h$, we can solve for the population of the holes (or equivalently the trions),
\begin{equation}
\frac{dn_{h,T}}{n_{h,T}^2} = - R_T \, dt
\quad \Rightarrow \quad
n_{h,T}(t) =
\frac{1}{n_{h,T}^{-1}(t_0) + R_T (t - t_0)} .
\label{eq:holes_trions}
\end{equation}

We observe that the hole (as well as the trion) population exhibits a slow power-law decay,
\[
n_{h,T}(t) \sim \frac{1}{R_T t},
\]
at late times. Plugging the solution for the holes and trions into the equation for the excitons, we obtain a closed-form solution for the exciton population,
\begin{equation}
n_X(t) = e^{-R_X t}
\int_{t_0}^{t}
\frac{2 R_T e^{R_X t'}}{\left[n_h^{-1}(t_0) + R_T (t' - t_0)\right]^2}
\, dt' ,
\label{eq:exciton_solution}
\end{equation}
which can be written explicitly in terms of the exponential integral function.

Considering the asymptotic behavior as $t \to \infty$, we find
\[
n_X(t) \sim \frac{1}{R_X R_T t^2}.
\]
This implies that the dominant contribution to the QPF at late times is expected to originate from the trions and exhibit a $t^{-1}$ decay in intensity. The power law behavior cannot be observed within the delay range of the experimental data. However, examining the fitted $n_X(t)$, $n_T(t)$, and their sum, confirms the asymptotic behavior and is summarized in Table \ref{tab:late_times}.

\begin{table}
	\centering

	\begin{tabular}{lccr}
		\\
		\hline
		Fluence ($\mu$J/cm$^2$) &  $n_X$ power & $n_T$ power & $n_X+n_T$ power\\
		\hline
		490 (high)&-2.07&-0.993&-1.013\\
		160 (medium)&-1.9930&-0.991&-1.001\\
		50 (low)&-1.974&-0.983&-0.991\\
		\hline
	\end{tabular}
    	\caption{Late times power law.
		Values for the late times ($40-50$~ps) power law exponents for excitons, trions, and total quasiparticle populations, evaluated for the fit results presented in Fig.3 and in Table 1.}
        	\label{tab:late_times}
\end{table}

\subsection{Rate equation electron and holes}
\begin{figure}
	\centering
	\includegraphics[width=0.75\textwidth]{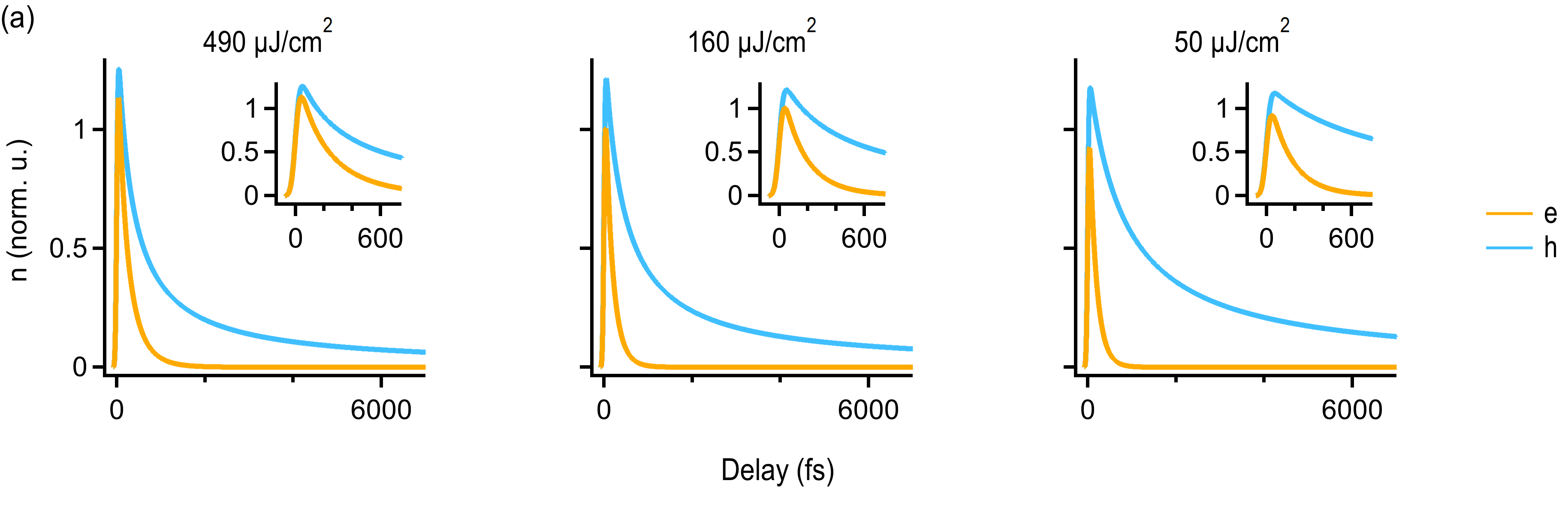} 
	\caption{Electrons and holes. Calculated electron (orange) and hole (light-blue) population for the QP fits of Fig.3 for the three fluences.
}
	\label{fig:s_e_and_h}
\end{figure}

The rate equation describes quasiparticle buildup (and decay), starting with CB electrons and VB holes. Figure.~\ref{fig:s_e_and_h} shows the electron and hole population for the fit of the rate equation presented in Fig.3(d). In all three fluences the electron population decays within a few hundred femtoseconds, as the quasiparticles appear, matching the observed rapid evacuation of the CB in Fig.2(e).

\bibliography{paper_bib_with_urls}

\clearpage